\input harvmac
\input epsf


\lref\rWitten{E. Witten, {\it String theory dynamics in various 
dimensions,} Nucl. Phys. B443 (1995) 85, hep-th 9503124.}
\lref\rBFSS{T. Banks, W. Fischler, S.H. Shenker and L. Susskind,
{\it M Theory as a Matrix Model: A Conjecture}, Phys. Rev. D55 
(1997) 112.}
\lref\rSusskind{L. Susskind, {\it Another Conjecture about M(atrix)
Theory,} hep-th 9704080.}
\lref\rTaylor{W. Taylor, {\it D-brane field theory on compact 
spaces,} Phys. Lett. B394 (1997) 283, 
hep-th 9611042.}
\lref\rDVV{L.Motl, {\it Proposals on Nonperturbative Superstring 
Interactions,}
hep-th 9701025\semi                  
T.Banks and N.Seiberg, {\it Strings from Matrices,}
Nucl. Phys. B497 (1997) 41, hep-th 9702187\semi
R.Dijkgraaf, E.Verlinde and H.Verlinde, {\it Matrix String Theory,}
Nucl. Phys. B500 (1997) 43, hep-th 9703030.}
\lref\rUsO{R. de Mello Koch and J.P. Rodrigues, {\it Classical 
integrability of
chiral $QCD_{2}$ and classical curves,} hep-th 9701138, to be 
published
Mod. Phys. Lett. A12 (1997) 2445.}
\lref\rUsT{R. de Mello Koch and J.P. Rodrigues, 
{\it The dynamics of classical chiral
$QCD_{2}$ currents,} hep-th 9708080.}
\lref\rHo{P.M.Ho and Y.S.Wu, {\it IIB duality and Longitudinal 
Membranes
in M(atrix) Theory,} hep-th 9703016.}
\lref\rGSW{M.B.Green, J.H.Schwarz and E.Witten, {\it Superstring 
Theory,}
Vol.1, CUP (1987).}
\lref\rCrem{E.Cremmer, H.L\"u, C.N.Pope and K.S.Stelle, {\it Spectrum-
generating
Symmetries from BPS Solitons,} hep-th 9707207.}
\lref\rBanks{T. Banks, {\it Matrix Theory,} hep-th 9710231\semi
D. Bigatti and L. Susskind, {\it Review of Matrix Theory,} 
hep-th 9712072.}
\lref\rSeibergLight{N. Seiberg, {\it Why is the Matrix Model 
Correct?},
Phys. Rev. Lett. 79 (1997) 3577, hep-th 9710009\semi
A. Sen, {\it D0-branes on $T^N$ and matrix theory"}, hep-th/9709220.}
\lref\rSussMatch{Susskind matching paper}
\lref\rSchwarz{J.H.Schwarz, {\it Lectures on Superstring
and M Theory Dualities,} Nucl.Phys.Proc.Suppl. 46 (1996) 
30, hep-th 9607201\semi 
J.H.Schwarz, {\it The power of M theory,} Phys.Lett. 
B367 (1996) 97, hep-th 9510086\semi
J.H.Schwarz, {\it An $SL(2,Z)$ multiplet of
type $IIB$ superstrings,} Phys. Lett. B360 (1995) 13; erratum ibid.
B364 (1995) 252; hep-th 9508143\semi
P.S. Aspinwall, {\it Some relationships between dualities in
string theory}, Nucl.Phys.
Proc.Suppl. 46 (1996) 30, hep-th 9508154.}
\lref\rPolchinski{J. Polchinski, {\it TASI Lectures on D-branes,} 
hep-th 9611050.}
\lref\rBO{M. Blau and M. O'Loughlin, {\it Aspects of U duality in 
Matrix Theory,} hep-th 9712047.}
\lref\rHT{C.M. Hull and P.K. Townsend, {\it The Unity of Superstring
Dualities,} Nucl.Phys.B438 (1995) 109, hep-th 9410167.}
\lref\rHull{C.M. Hull, {\it Matrix Theory, $U$ duality and Toroidal
Compactifications of $M$ theory,} hep-th 9711179.}
\lref\rBP{C.M. Hull, {\it U Duality and BPS Spectrum of Super Yang-
Mills
and M-theory,} hep-th 9712075\semi
N.A. Obers, B. Pioline and E. Rabinovicci, {\it M-theory and U-duality
on $T^{d}$ with Gauge Backgrounds,} hep-th 9712084.}
\lref\rJulia{C.Hull and B. Julia, in preparation}
\lref\rHV{F. Hacquebord and H. Verlinde, {\it Duality Symmetries of 
N=4
super Yang-Mills theory on $T^{3}$,} Nucl. Phys. B508 (1997) 609, hep-
th
9707179.}
\lref\rKPV{H. Verlinde, {\it Interpretation of the Large N Loop
Equation,} hep-th 9705029\semi
S. Gukov, I.R. Klebanov and A.M. Polyakov, {\it Dynamics of
$(n,1)$ strings,} hep-th 9711112.}
\lref\rPLT{C. Klimcik and P. Severa, {\it Non-Abelian Momentum Winding
Exchange,} Phys. Lett. B383 (1996) 281, hep-th 9605212.}
\lref\rUsThree{R.de Mello Koch and J.P. Rodrigues, {\it Duality
and Symmetries of the Equations of motion,} hep-th 9709089v1.}
\lref\rKut{S. Elitzur et al, {Algebraic Aspects of Matrix Theory on 
$T^d$,}
Nucl. Phys. B509 (1998) 122, hep-th 9707217.}


\Title{ \vbox {\baselineskip 12pt\hbox{CNLS-98-01}
\hbox{BROWN-HET-1108}  \hbox{January 1998}  }}
{\vbox {\centerline{Duality and Light Cone Symmetries of the}
        \centerline{Equations of Motion}
}}

\smallskip
\centerline{Robert de Mello Koch}
\smallskip
\centerline{\it Physics Department and Centre for Nonlinear Studies}
\centerline{\it University of the Witwatersrand, Wits 2050, South 
Africa}
\bigskip
\centerline{\it and}
\bigskip
\centerline{Jo\~ao P. Rodrigues\foot{Work supported in part by the 
Department 
of Energy under contract DE-FG02-91ER40688-A.}}
\smallskip
\centerline{\it Physics Department and Centre for Nonlinear 
Studies\foot{
Permanent Address}}
\centerline{\it University of the Witwatersrand, Wits 2050, South 
Africa}
\smallskip
\centerline{\it and}
\centerline{\it Department of Physics}
\centerline{\it Brown University, Providence, RI 02912}
\bigskip

\medskip

\noindent
The matrix theory description of the discrete light cone quantization
of $M$ theory on a $T^{2}$ is studied. 
In terms of its super Yang-Mills description, we identify symmetries 
of the
equations of motion corresponding to independent rescalings of one of
the world sheet light cone coordinates, which 
show how the $S$ duality 
of Type IIB
string theory is realized as a Nahm-type transformation.
In the $M$ theory description this corresponds to a simple $9-11$ 
flip.


\Date{}


Dualities play a central role in our current understanding of string
and supersymmetric gauge theories \rWitten. Since supergravity is the 
low energy limit of string theory, a proper understanding of the 
symmetries 
present in these supergravity theories is bound to shed light on 
string 
theory. This has in fact proved to be a very useful approach, 
culminating in
a comprehensive classification of $U$ duality symmetries acting in 
various
compactified and uncompactified string theories \rHT. 

Recently, Susskind put forward a remarkable conjecture 
\rSusskind\ linking the discrete
light cone quantization of $M$ theory on $T^{d}\times R^{10-d}\times 
S^{1}$
to a $U(N)$ super Yang-Mills theory on $T^{d}\times R$ for finite $N$.
(This was preceded by the m(atrix) theory conjecture of \rBFSS; a
derivation of both conjectures as well as the relationship between 
the two
has been suggested in \rSeibergLight.) For 
$d<4$, the super Yang-Mills theory
provides a complete description whilst for $d>3$ it is only valid as 
a low
energy description. Based on our experience with string theory, it is
natural to expect that a great deal of information about $M$ theory is
contained in these super Yang-Mills theories, in much the same way 
that
supergravity provides important insights into string theory \rHull. 
Conversely,
since super Yang-Mills theories are interesting in their own right, 
it is
also interesting to see what can be learnt about Yang-Mills theories 
from
$M$ theory.

Super Yang-Mills theories have the well known property that all 
dependence on
the coupling constant $g_{YM}$ can be factored out of the Lagrangian 
density. Since in general $g_{YM}$ is dimensionful, it immediatly 
follows that
the action will scale homogeneously under a worldvolume scale 
transformation
(i.e., by multiplication by a $d$ dependent
power of the dimensionless parameter of the tranformation).
A scale tranformation will also change the volume of the 
compactification 
torii and therefore scale transformations provide a possible 
framework in which
one can study the dependence of coupling constants on 
compactification radii. Since the action transforms homogeneously, it 
follows that these transformations are symmetries of the classical 
equations of motion but they are {\it not} in
general symmetries of the action. 

With this in mind, we concentrate in this letter on the $d=1$ case, 
and try to see what 
insights can be gained from the description of $M$ theory in terms of 
a super Yang-Mills theory. This theory corresponds to $M$ theory on a 
$T^{2}$, or to a two dimensional super Yang-Mills theory on an 
$S^{1}$. 
Introducing world-sheet light cone coordinates $\sigma^{\pm}$, we 
point out
symmetries of the equations of motion that allow one to rescale 
$\sigma^+$
and $\sigma^-$ {\it independently}. We will refer to these 
independent 
light cone rescalings as "lico" scale transformations. 

We will then use the $\sigma^{-}$ "lico" scale tranformation to 
establish the existence of a Nahm-type transformation
mixing the rank of the gauge group and the
electric flux. This can be identified with $S$ duality in Type IIB 
string
theory
and with a flip of the two cycles of the $T^{2}$ in $M$ theory. As we 
show,
flipping two cycles in $M$ theory is a trivial operation, so that the 
link
between the discrete light cone quantization of $M$ theory and super
Yang-Mills theory explains the existence of Nahm transformations. 
Similar
points of view have recently appeared in \rBO, \rBP; the relevance of 
Nahm transformations was first pointed out in \rHV.

Our starting point is the M(atrix) description of
$M$ theory \rBFSS. The relevant Lagrangian, written in 
units where the string scale $l_{s}=1$, is \rBFSS,
\eqn\MatrixLag
{L={1\over 2g}tr\Big[ \dot{X}^{I}\dot{X}^{I}
+2i\bar{\theta}\dot{\theta}
+{1\over 2}\big[X^{I},X^{J}\big]^{2}
-2\bar{\theta}\gamma_{I}\big[\theta ,
X^{I}\big]\Big] , \qquad  l_{p}^{3} = R_{11}l_{s}^{2}. }

The relationship between the Planck scale and string scale identifies 
the
compactification radius $R_{11}$ with the type IIA string coupling 
constant
$g$. From this Lagrangian a Hamiltonian $H$ follows in the usual 
manner. As $R_{11} \to 0$, the diagonal entries of the $N \times N$ 
hermitean 
matrices $X^I,I=1,...,9$ are associated with the coordinates of $N$ D 
particles in type IIA string theory, in the sector with total 
momentum 
$P=N/R_{11}$. In this limit and for finite N, a suitable redefinition 
of 
parameters can be made \rSeibergLight\  that maps the above theory to 
the M(atrix) description of the DLCQ of M theory \rSusskind\
in the sector $P^{+}= N/R$. $R$ is a lightlike compactification radius
\rSeibergLight.

To compactify the theory \MatrixLag\ on a transverse circle of radius 
$R_{9}$, instead 
of choosing to identify the coordinate $X^9$ with a covariant 
derivative
 \rDVV, we take the more pedestrian approach followed originally in 
\rBFSS\ and \rTaylor\  of tessellating the $X^9$ coordinate by 
requiring 

\eqn\Taylor
{\eqalign{X^9_{n,n} &= X^9_{n-1,n-1} + 2 \pi R_9 , 
\qquad X^9_{n,m} = X^9_{n-1,m-1} \cr  
X^I_{n,n} &= X^I_{n-1,n-1} ,\quad  X^I_{n,m} = X^I_{n-1,m-1},
\quad I=1,...,8. 
}}

where each $X^I_{n,m}$ is a block matrix. After identifying 
$X^I_{m,n} = X^I_{m-n}=X^{I\dagger}_{n-m}$, 
we find \rTaylor\

\eqn\BosonicPiece
{\eqalign{L&={1\over 2 R_{11}}tr\sum_{n=0}^{\infty}\Big[\sum_{I=1}^{9}
\dot{X}^{I}_{n}
\dot{X}^{I}_{-n}-\sum_{J=1}^{8}S_{n}^{J}(S_{n}^{J})^{\dagger}-
{1\over 2}\sum_{J,K=1}^{8}T_{n}^{JK}
(T_{n}^{JK})^{\dagger}\Big],\cr
S_{n}^{J}&=\sum_{q}\big[ X_{q}^{9},X_{n-q}^{J}\big]-2\pi
R_{9}nX_{n}^{J},\cr
T_{n}^{JK}&=\sum_{q}\big[ X_{q}^{J},X_{n-q}^{K}\big] .}}

The interpretation of the fields appearing in the above Lagrangian is 
that the $X_{0}^{I},\quad I = 1,...,9$ describe a configuration of 
$D0$ 
branes whilst the $X_{n}^{I}, \quad I=1,...,8$ describe open strings 
stretching between these $D$ particles which wind $n$ times around 
the 
compact dimension $x^{9}$. This is easily verified by means of a 
simple 
semi-classical computation \rTaylor.

The (t-dependent) solution

\eqn\beforeDONine
{X^9_0 = {{R_{11}} \over {R_9}} {n \over N} t \quad 1_{N\times N},}

with $1_{N\times N}$ the $N\times N$ unit matrix, carries KK momentum 
$n/R_{9}$ around 
$R_9$. 
 
Introducing the expansions 

\eqn\FourierSeries
{A^{9}={1\over2\pi}\sum_{n}e^{inx/\bar{R}_9}X_{n}^{9}\quad
X^{I}={1\over 2\pi}\sum_{n}e^{inx/\bar{R}_9}X_{n}^{I},
\quad I=1,...,8}

we find the action of the $T$ dual, two dimensional super Yang Mills 
theory

\eqn\DIStrings
{\eqalign{S&=\int_0^{\bar{R}_9} {dtdx^{9} {l_s}^2 \over 2
R_{11}{R_9}^{-1}}
tr\Big[ \dot{X}^{I}\dot{X}^{I}
+\dot{A}^{9}\dot{A}^{9}-
(\partial_{9}X^{I}-i\big[A^{9},X^{I}\big])^{2}
+{1\over 2}\big[X^{I},X^{J}\big]^{2}\cr
&+2i\bar{\theta}\gamma^{\mu}D_{\mu}\theta
+2\bar{\theta}\gamma_{I}\big[X^{I},\theta\big]\Big], 
\qquad  P={N \over R_{11}} }}

of $D1$ branes \rTaylor\ \rBanks. As usual, 
$\bar{R}_9={l_{s}^{2}\over R_{9}}$. We have reinstated the fermions 
and 
$l_s$ for fields with the dimensions appropriate to a SYM description.

The solution \beforeDONine\ translates into an electric flux on the 
worldsheet of the $D$ string in the $T$ dual picture. The electric 
flux is 
quantized in units of $\bar{R}_{9}$

\eqn\DFlux
{  E = {\partial L \over \partial {\dot{A}^9}} = 
{\dot{A}^9 {l_s}^4 \over R_{11}}
     = {\bar{R}_9 {n \over N} }1_{N\times N}, \qquad  \dot{A}^9 = 
{R_{11} \over R_9} {n \over N} {1 \over l_s^2}1_{N\times N} . }

Its SYM energy is given by

\eqn\ESYM
{ E_{SYM} = {n^2 \over 2 N} {R_{11} \over R_9^2 }}

The M theoretic nine dimensional mass of this non-threshold BPS state 
in
a $N D1$ background is:

\eqn\DSchwarz
{ m^{2}= 
{N^2 \over R_{11}^2} + {n^2 \over R_9^2} =
{ 1 \over R_{9}^2} \Big[\Big( {R_{9} \over R_{11} }\Big)^2 
N^{2}+n^2\Big].} 

This can be argued as follows. Certainly

\eqn\MassOne
 {        E_{SYM} = \lim_{P \to \infty}
         \Bigl[   \sqrt{ P^2 + 2 P E_{SYM} } - P \Bigr] ,} 

and if in addition to $R_{11} \to 0$ we let $l_s \to 0$ with 
$R_{11}/l_p^2$ and $R_9 / l_p$ constant \rSeibergLight, then the 
string oscillators decouple. The energy $E_{SYM}$ as well as the 
parameters 
$g_{YM}^2 = R_{11}/(R_9 l_s^2)$ and $\bar{R}_9$ of the action 
\DIStrings\ 
remain finite.

We will interpret equation \DSchwarz\ in terms of strings winding the 
cycle
of length $\bar{R}_{9}$. In this case, this formula matches the mass 
of a 
string carrying $N$ units of $RR$ charge and $n$ units of $NS-NS$ 
charge, in 
Type IIB string theory \rSchwarz. 
In addition, the Type IIB string 
coupling is $g=R_{11}/R_{9}$, in agreement with the identification 
of \DIStrings\ as a low energy effective action of $D1$ branes
\rPolchinski.

It is well known that Type IIB string theory has an $S$ duality which 
acts
by inverting the string coupling constant and exchanging $NS-NS$ and 
$RR$
charged states. We are interested in understanding how this duality 
arises 
in the super Yang-Mills description we are considering here. Towards 
this
end, we rewrite this action in light cone coordinates
$\sigma^{\pm}={1\over\sqrt{2}}(x^{0}\pm x^{9})$
and in the lightcone gauge $A_{-}=0$

\eqn\DStrings
{\eqalign{S&={l_s^2 \over 2 R_{11} R_9^{-1}}\int d\sigma^{+}
\int^{\tilde{R}}_{0}
d\sigma^{-}tr\Big[
2\partial_{+}X^{I}\partial_{-}X^{I}
+i\sqrt{2}\theta^{T}_{R}\partial_{+}
\theta_{R}+i\sqrt{2}\theta^{T}_{L}\partial_{-}\theta_{L}
+J^{+}A_{+}-\cr
&-2(\partial_{-}A_{+})^{2}+{1\over 2}\big[ X^{I},X^{J}\big]^{2}+
2\theta^{T}_{L}\gamma_{I}\big[ X^{I},\theta_{R}\big]\Big],\cr
J^{+}&=2\theta_{R}^{T}\theta_{R}
+2i\big[ X^{I},\partial_{-}X^{I}\big], }}

The limits of integration in this equation are to be understood as 
follows: 
due to the two-dimensional Lorentz invariance of the Lagrangian 
density,
one can perform a large boost with parameter 

\eqn\Boost
{\beta = { 
{\tilde{R}} \over
{\sqrt{\tilde{R}^2+2\bar{R}_9^2}}} \approx  
1 - {{\bar{R}_9^2} \over {\tilde{R}^2}} } 

which maps the range of integration in equation \DIStrings\ to the
lightlike $S^1$ compactification of equation \DStrings, in the
limit $\tilde{R}>>\bar{R_9}$. This argument is essentially similar 
to the space-time argument of \rSeibergLight.  

The Euler-Lagrange equations
which follow from \DStrings\ are

\eqn\BBeqnsofmotion
{\eqalign{\partial_{-}^{2}A_{+}&=-{1\over 4}J^{+},\cr
\partial_{-}\theta_{L}&={i\gamma_{I}\over\sqrt{2}}
\big[X^{I},\theta_{R}\big],\cr
4\partial_{+}\partial_{-}X^{I}&=2i\Big(2\big[\partial_{-
}X^{I},A_{+}\big]
+\big[ X^{I},\partial_{-}A_{+}\big]\Big)+2\big[ X^{J},\big[X^{I},X^{J}
\big]\big]-2(\gamma_{I})_{\alpha\beta}\{\theta_{L}^{T\alpha},
\theta_{R}^{\beta}\},\cr
i\partial_{+}\theta_{R}&={1\over\sqrt{2}}\big[A_{+},\theta_{R}\big]
-{1\over \sqrt{2}}\big[ X^{I},\theta_{L}\big]\gamma_{I}.}}

Using the above equations of motion, it is not difficult to show that
purely bosonic solutions with $\theta_{L,R}=0$ exist.
For this set of solutions and in the case when the $X^{I}$'s are
mutually commuting, one obtains the equation

\eqn\DTDXDXA
{i\partial_{+}\partial_{-}^{2}A_{+}=
{1\over 4}\big[\partial_{-}^{2}A_{+},A_{+}\big].}

This equation has been studied extensively in \rUsO, \rUsT. It is of 
the 
Lax form and for $SU(2)$ has a rich structure including multi-soliton 
solutions, an 
auto-B\"acklund transformation and a non-linear superposition 
principle 
\rUsT. It also has a diffeomorphism invariance in the $\sigma^{-}$ 
coordinate which is not a symmetry of the action, as first pointed 
out in
\rUsO. 

It is not difficult to verify that the set of transformations of the 
form 
$\sigma^{-}\to a\sigma^{-}, \quad \sigma^{+}\to \sigma^{+}$ with $a$ 
a 
constant are symmetries of \BBeqnsofmotion, with the fields scaling as

\eqn\TransformFields
{\eqalign{A_{+}(\sigma^{+},\sigma^{-})&\to 
A'_{+}(\sigma^{+},\sigma^{-})=A_{+}(\sigma^{+},a\sigma^{-}),\cr
X^{I}(\sigma^{+},\sigma^{-})&\to X'^{I}(\sigma^{+},\sigma^{-})=
\sqrt{a}X^{I}(\sigma^{+},a\sigma^{-}),\cr
\theta_{R}(\sigma^{+},\sigma^{-})&\to\theta'_{R}
(\sigma^{+},\sigma^{-})=a\theta_{R}(\sigma^{+},a\sigma^{-}),
\cr
\theta_{L}(\sigma^{+},\sigma^{-})&\to\theta'_{L}
(\sigma^{+},\sigma^{-})=\sqrt{a}
\theta_{L}(\sigma^{+},a\sigma^{-}),\cr
J^{+}(\sigma^{+},\sigma^{-})&\to J'^{+}(\sigma^{+},\sigma^{-})
=a^{2}J^{+}(\sigma^{+},a\sigma^{-}).}}

Under the symmetry \TransformFields\ the Lagrangian transforms as

\eqn\TransLag
{{1\over 2g\bar{R_9}}\int^{\tilde{R}}_{0}
d\sigma^{-} L(A_{+},J^{+},X^{I},\theta_{R},\theta_{L})\to
{1\over 2ga\bar{R_9}}\int^{\tilde{R}/a}_{0}
d\sigma^{-} L(A_{+},J^{+},X^{I},\theta_{R},\theta_{L})}

explicitly demonstrating that \TransformFields\ is not a symmetry of 
the action: indeed the action is a homogeneous function under the
scaling \TransformFields. This is independent of the choice of 
gauge. 

These tranformations can be understood in a more fundamental way
by verifying that the theory has two independent length scales for the
$\sigma^{+}$ and $\sigma^{-}$ coordinates. Introducing   

\eqn\DimOfCoords
{\big[ \sigma^{+}\big]=l_{+}\quad \big[ \sigma^{-}\big] =l_{-},}

and requiring that all of the terms appearing in the Lagrangian have 
the
same dimension, leads to the following dimensions for the different
fields entering into the action

\eqn\DimOfFields
{\eqalign{\big[ A_{+} \big]=l_{-}^{0}l_{+}^{-1}\cr
\big[ A_{-} \big]=l_{-}^{-1}l_{+}^{0}\cr
\big[ X^{I}\big]=l_{-}^{-{1\over 2}}l_{+}^{-{1\over 2}}\cr
\big[ \theta_{R}\big]=l_{-}^{-1}l_{+}^{-{1\over 2}}\cr
\big[ \theta_{L}\big]=l_{-}^{-{1\over 2}}l_{+}^{-1}.}}

The symmetries of \TransformFields\ are then seen to correspond to
scale transformations in the $\sigma^{-}$ coordinate.
It follows that a symmetry rescaling only the $\sigma^{+}$ coordinate
is also present, or more precisely, the equations of motion have 
a two parameter set of light cone symmetries  
$\sigma^{-}\to a\sigma^{-}$ and $\sigma^{+}\to b\sigma^{+}$
with the corresponding field transformations respecting the light cone
dimensions \DimOfFields\ .

These tranformations can also be understood as compositions of the 
usual scale tranformation of both coordinates with a Lorentz boost. 
This alternative 
interpretation allows
an easy generalization of higher dimensional world volumes.
As indicated earlier, to stress the light cone nature of the
above scale transformations, we will refer to them as "lico scale 
transformations".
\foot{We remark that it has recently been pointed out that the 
equations
of motion of type $IIB$ supergravity in the bosonic sector
have a (field) rescaling symmetry not present at the level of the 
action
\rCrem.}

We will now specialize to the parameter
\foot{We set $b=1$, ie, we only consider $\sigma^{-}$ rescalings in 
this letter.}

\eqn\Parameter
{ a = {l_s^2 \over l_s^{'2}} {{R_{11}} \over {R}_{9}}.}

anticipating a different string scale $l_s'$ after the transformation.
The relationship between the two different string lenghts will be 
discussed 
later. 

After boosting back, the action \DIStrings\ maps into 

\eqn\DFStrings
{\eqalign{S&=\int_0^{\bar{R}_{11}} {{dtdx^{9} l_s^{'2} R_9^2 }\over 
2 R_{11}^2}
tr\Big[ \dot{X}^{I}\dot{X}^{I}
+\dot{A}^{9}\dot{A}^{9}-
(\partial_{9}X^{I}-i\big[A^{9},X^{I}\big])^{2}
+{1\over 2}\big[X^{I},X^{J}\big]^{2}\cr
&+2i\bar{\theta}\gamma^{\mu}D_{\mu}\theta
+2\bar{\theta}\gamma_{I}\big[X^{I},\theta\big]\Big]
\qquad P = {N \over R_{11}} .    }}

In the above $\bar{R}_{11} = l_s^{'2} / R_{11}$. Let us consider
now the state with flux quantized in units of $\bar{R}_{11}$:

\eqn\DTFlux
{  E = {\partial L \over \partial {\dot{A}^9}} = 
{\dot{A}^9 {l_s}^{'4} R_9^2 \over R_{11}^3}
     = {\bar{R}_{11} {n \over N} }1_{N\times N}, \qquad  \dot{A}^9 = 
{R_{11}^2 \over R_9^2} {n \over N} {1 \over l_s^{'2}}1_{N\times N} . }

Its SYM energy is

\eqn\ESYMprime
{ E'_{SYM} = {n^2 \over 2 N} {R_{11} \over R_9^2 }}

and
one finds that the M-theoretic nine dimensional mass associated with 
the 
state is { \sl identical\/} to that of \DSchwarz:

\eqn\FSchwarz
{ m^{2}= 
{N^2 \over R_{11}^2} + { n^2 \over R_9^2} =
{ 1 \over R_{11}^2} \Big[ N^2 + \Big({R_{11} \over R_{9} }\Big)^2
n^2\Big].}

This mass must now be interpreted 
in terms of strings winding $\bar{R}_{11}$. In this
case, we find that we have the mass of a state with $N$ units of $NS-
NS$
charge and $n$ units of $RR$ charge. In addition, we can read off the 
string
coupling as $R_{9}/R_{11}$, which has been inverted. In other words,
in this implementation of the symmetry one obtains a description of 
the
same state from different points of view.

Clearly then, the lico scale transformation implements a 
transformation involving a "$9-11$ flip"; 
in terms of the Yang-Mills
picture, we
have swopped the role of electric flux with the rank $N$ of the gauge 
group which
is reminiscent of a Nahm transformation. This has been the subject of 
a 
number of recent studies \rKPV.

There are a few important points which should now be addressed. First,
since the string tension remains invariant under a $T$ duality,
the "$9-11$ flip " must be responsible for the change in string 
tension
associated with the duality exchanging $RR$ 
charges and
$NS-NS$ charges.  
The easiest way to 
see this is 
to note that in Planck units one is simply exchanging
$R_9$ and $R_{11}$.  Before the flip, the relationship between the 
string
scale and the Planck length is $l_{p}^{3}=R_{11}l_{s}^{2}$. However,
after the flip we have $l_{p}^{3}=R_{9}l_{s}^{'2}$. Therefore 

\eqn\StringTensions
{l_{s}^{'-2}={R_{9}\over R_{11}}l_{s}^{-2}=g^{-1}l_{s}^{-2}.}

Clearly, with this idenfication, equation \DFStrings\  is simply a
rewritting of the action \DIStrings, with the parameter of 
the "lico" scale transformation $a=1$. Remarkably, the correct
action is obtained independently of the relationship between the
two string lengths, providing consistency to the $SYM$ formalism
and in particular to the implementation of our symmetry.

Let us now analyze the above tranformation on the parameters of the
theory, in constant string units. One straightforwardly obtains

\eqn\Map
{\eqalign{ \bar{R}_9={l_s^2 \over R_9} &\to {l_s^{'2} \over R_9'}=
 {{l_s}^{'2} \over R_{11}}={l_s^2 \over R_9}=\bar{R}_9\cr
 g_{YM}^2 = {g \over l_s^2} = {R_{11} \over R_9 l_s^2} &\to
 {R_9 \over R_{11} l_s^{'2}} = {R_9^2 \over R_{11}^2 l_s^2 }= 
g_{YM}^2 
 \Big({R_9 \over R_{11}}\Big)^3\cr
g={R_{11} \over R_9} &\to g'= {1 \over g}. }}

Can one implement the transformation in this form using the "lico"
symmetries? The answer is yes: consider the "lico" scale 
transformation
with parameter $a=R_{11}/R_9$. One obtains 

\eqn\DDStrings
{\eqalign{S&=\int_0^{l_s^2 \over {R_{11}}} 
{{dtdx^{9} {l_s}^2 R_9^2 }\over 
2 R_{11}^2}
tr\Big[ \dot{X}^{I}\dot{X}^{I}
+\dot{A}^{9}\dot{A}^{9}-
(\partial_{9}X^{I}-i\big[A^{9},X^{I}\big])^{2}
+{1\over 2}\big[X^{I},X^{J}\big]^{2}\cr
&+2i\bar{\theta}\gamma^{\mu}D_{\mu}\theta
+2\bar{\theta}\gamma_{I}\big[X^{I},\theta\big]\Big]
\qquad P = {N \over R_{11}} .    }}

Since the duality \Map\  requires $\bar{R_9}$ to remain invariant,
in order to be able to compare with \DIStrings, 
we simply relabel $R_9 \leftrightarrow R_{11}$ in \DDStrings, and we 
obtain

\eqn\DDFStrings
{\eqalign{S&=\int_0^{l_s^2 \over {R_{9}}} 
{{dtdx^{9} {l_s}^2 R_{11}^2 }\over 
2 R_9^2}
tr\Big[ \dot{X}^{I}\dot{X}^{I}
+\dot{A}^{9}\dot{A}^{9}-
(\partial_{9}X^{I}-i\big[A^{9},X^{I}\big])^{2}
+{1\over 2}\big[X^{I},X^{J}\big]^{2}\cr
&+2i\bar{\theta}\gamma^{\mu}D_{\mu}\theta
+2\bar{\theta}\gamma_{I}\big[X^{I},\theta\big]\Big]
\qquad P = {N \over R_{9}} ,    }}

which precisely reproduces the change in the SYM coupling predicted by
\Map. Calculating the SYM energy of the electric flux state of 
\DDFStrings\ 
quantized in units of $\bar{R}_9$, as done previously, we obtain: 

\eqn\ESYMp
{ E'_{SYM} = {n^2 \over 2 N} {R_{9} \over R_{11}^2 }}

Comparison with \ESYM\  shows explicitly that the two states have been
obtained from each other by the exchange $R_9 \leftrightarrow R_{11}$.

One should notice that the rescalings \rSeibergLight\  taking 
the theory \DIStrings\ in the sector with momentum 
$P=N/R_{11}$ and compactification radius $R_9$ to the matrix DLCQ of M
theory in the sector with $P^+=N/R$ ($R$ is light like) and 
corresponding
compactification radius $\tilde{R}_9$ (say) leave 
$E_{SYM}$ {\it invariant}. Therefore, after the duality transformation
one would arrive at \ESYMp\ with $R_{11}$ and $R_9$ 
replaced by $R$ and $\tilde{R}_9$, respectively. Therefore in the
matrix DLCQ of M theory, the above duality exchanges the null
circle $R$ with the space-like circle $\tilde{R}_9$. There is evidence
that apparently this is to be expected \rJulia\ (as referred to
in \rHull\ and \rBO ). This is also related to the need to consider
the range of integration in \DStrings\ as that of a light like circle
in a static gauge,  
although this aspect deserves further study outside the scope of this
letter. 

If one associates for large $P$ a nine dimensional M-theoretic mass 
to the electric flux state, as done previously, one obtains

\eqn\FinalMass
{m^2  = {n^2 \over R_{11}^2 }+  {N^2 \over R_9^2}  .     } 

One sees explicitly the roles of $R_9$ and $R_{11}$ (or equivalently
$n$ and $N$ exchanged. In this case duality is implemented by changing
the mass of the state \DSchwarz.

This duality has recently been discussed
in \rBO\  \rBP, where it was designated N-duality. It was shown
that N-duality is essential to the extension of the U duality group 
$E_d$
\rKut
to the full $E_{d+1}$ U-duality group in the YM description of
m(atrix) theory on $T^{d+1}$. 

In summary, the "lico" scale transformation has three very different
interpretations: In the Yang-Mills theory, it plays the role of a Nahm
type of transformation swopping the roles of the electric flux and the
rank of the gauge group. In the Type IIB theory, it plays the role of 
an
$S$ duality taking us from a theory with coupling $g=R_{11}/R_{9}$ and
string tension $l_{s}^{-2}$, to a new theory with coupling $g^{-1}$ 
and
string tension $g^{-1}l_{s}^{-2}$. In $M$ theory, it is a trivial
transformation corresponding to swopping the two cycles of the 
$T^{2}$ on
which the theory is compactified.

Finally, we should remark that "9-11 flip" discussed here is not 
related
to the "9-11 flip" of references \rDVV. Indeed, inspection of 
equations
\DFStrings\  and \DDFStrings\ shows that in both cases the radius
of the SYM $S^1$ is {\it not} the direction associated with the 
($IIA$) string coupling, ie the direction chosen to relate the 
Planck and string lengths.

{\it Note Added} This work is an expanded and clarified version of our
previous results reported in \rUsThree.

{\it Acknowledgements} R.de M.K. is grateful to the Centre for 
Nonlinear
Studies at Wits for a travel grant allowing him to participate in the
"CERN Workshop on Branes and Non-perturbative Dualities", and to the
organizers and lecturers of the workshop for providing such a 
stimulating
forum for discussion. J.P.R. thanks the High Energy Theory Group at 
Brown for
their hospitality. This research is partially supported
by the FRD under grant number GUN-2034479.

\listrefs
\vfill\eject
\bye